\documentstyle[psfig,12pt]{article}

\def\Mpc{\,{\rm Mpc}}

\def\eV{\,{\rm eV}}

\def\gcm2{\,{\rm g\,cm}^{-2}}

\def\g{$\gamma$}

\def\la{\mathrel{\mathpalette\fun <}}

\def\fun#1#2{\lower3.6pt\vbox{\baselineskip0pt\lineskip.9pt
  \ialign{$\mathsurround=0pt#1\hfil##\hfil$\crcr#2\crcr\sim\crcr}}}
\begin{document}
\pagestyle{empty}
\begin{center}
\vspace{1.5cm}
{\Large \bf A BREAK IN THE HIGHEST}\\
\bigskip
{\Large \bf ENERGY COSMIC RAY SPECTRUM:}\\
\bigskip
{\Large \bf A SIGNATURE OF NEW PHYSICS?}

\vspace{.3in}

G.~Sigl$^{a,b}$, S.~Lee$^{a,b}$, D.~N.~Schramm$^{a,b}$ and
P.~Bhattacharjee$^c$\\

\vspace{0.2in}

{\it $^a$Department of Astronomy \& Astrophysics\\
Enrico Fermi Institute, The University of Chicago, Chicago, IL~~60637-1433}\\

\vspace{0.1in}

{\it $^b$NASA/Fermilab Astrophysics Center\\
Fermi National Accelerator Laboratory, Batavia, IL~~60510-0500}\\

\vspace{0.1in}

{\it $^c$Indian Institute of Astrophysics\\
Sarjapur Road, Koramangala, Bangalore 560 034, INDIA}\\

\end{center}

\vspace{0.3in}

\centerline{\bf ABSTRACT}
\medskip
Recent experimental data from the Fly's Eye and the Akeno array
seem to indicate significant structure in the ultrahigh energy
cosmic ray spectrum
above $10^{18}\eV$. A statistically significant dip has been
established at about $5\times10^{18}\eV$. In addition, each
experiment observed a different superhigh energy event above
$10^{20}\eV$ separated from the rest of the data by
about half a decade in energy. In this article we discuss what
this implies for the existence or non-existence of the
``Greisen-Zatsepin-Kuz'min cutoff'', a long lasting and still open
question in cosmic ray physics. This cutoff, caused by energy
losses in the cosmic microwave background, is predicted to occur
at a few times $10^{19}\eV$ if cosmic rays are produced by
shock acceleration of lower energy particles at extragalactic
distances. We show that from the
spectral point of view, sources nearer than a few $\Mpc$ are still
consistent with the data at the $1\sigma$ level, provided these
sources accelerate particles beyond $3\times10^{20}\eV$.
However, persistence of the apparent gap in the existing data
at the level of a 4 times higher total exposure would
rule out a wide
range of acceleration models at $98\%$ C.L., whether
they rely on nearby or extragalactic sources. This might hint to
the existence of a ``top down'' mechanism which produces an
additional hard component of
ultrahigh energy particles directly, say, by decay from some
higher energy scale in contrast to
bottom up acceleration of charged particles. In this
scenario a cutoff
followed by a pronounced spectral flattening and possibly even a
gap could naturally be formed.

\newpage
\pagestyle{plain}
\setcounter{page}{1}
\section{Introduction}
For almost thirty years it has been
clear that the cosmic microwave background (CMB) has profound
implications for the astrophysics of ultrahigh energy cosmic
rays (UHE CR). Most notably, nucleons are subject to photopion
losses on the CMB which lead to a steep drop in the interaction length
at the threshold for this process at about $6\times10^{19}\eV$. This effect
is known as the Greisen-Zatsepin-Kuz'min (GZK)
effect~\cite{Greisen,Stecker1}. For heavy nuclei the giant dipole
resonance which leads to photodisintegration produces a similar
effect at about $10^{19}\eV$~\cite{Stecker2}.
One of the major unresolved questions in
cosmic ray physics is the existence or non-existence of a cutoff
in the UHE CR spectrum below $10^{20}\eV$ which could be
attributed to these effects if the sources are further away than a
few $\Mpc$.

The interest in this question has renewed since recently events
with energies above the GZK cutoff have been
detected~\cite{Watson,Efimov,Egorov,Bird1,Bird2,Agasa1,Agasa2}.
Most strikingly, both the Fly's Eye
experiment~\cite{Bird1,Bird2} and the Akeno
array~\cite{Agasa1,Agasa2}
detected a different superhigh energy event significantly beyond
$10^{20}\eV$ as well as an apparent gap of about half a decade
in energy between the highest and second highest events. This
led to a vigorous discussion on the nature and
origin of these particles~\cite{Biermann,Elbert,Sigl,Halzen}.
In this article we show that the structure of the high energy end
of the UHE CR spectrum has the potential to provide powerful constraints
on a wide class of models for these extraordinary
particles in the near future. The options discussed in
the literature can be divided into two categories.

In ``bottom-up'' scenarios charged baryonic particles are
accelerated to the relevant ultrahigh energies. This could, for
example, be
achieved by ordinary first order Fermi acceleration
at astrophysical shocks~\cite{Hillas} or by linear acceleration
in electric fields as they could arise for instance in magnetic
reconnection events~\cite{Colgate}. The resulting injection
spectrum  of the charged primaries at the source is typically a
power law in energy $E$, $j_{\rm inj}(E)\propto E^{-q}$. In the
case of reconnection acceleration there is no clear-cut
prediction for the power law index $q$, but in case of shock
acceleration it satisfies $q\geq2$. We will refer to this latter
case in what we call conventional bottom-up acceleration
scenarios in the following. Secondary neutral particles
like \g-rays and neutrinos are only produced by primary
interactions in these scenarios~\cite{Yoshida1}.

In top-down scenarios the primary particles which can be
charged or neutral are produced at ultrahigh energies in the
first place, typically by quantum mechanical decay of
supermassive elementary ``X'' particles related to Grand Unified
Theories (GUT's).
Sources of such particles today could be topological defects
(TD's) left over from early universe phase
transitions caused by spontaneous breaking of symmetries
underlying these GUT's~\cite{Bh1}. Generic features of these
scenarios are injection spectra considerably harder (i.e.
flatter) than in case
of bottom-up acceleration and a dominance of \g-rays in the
X particle decay products\cite{Aharon}. Even monoenergetic
particle injection beyond the GZK cutoff can lead to rather hard
spectra above the GZK cutoff~\cite{Wald}.

The distinction between these scenarios is closely related to
the existence or non-existence of the GZK cutoff in the form of
a break in the spectrum. In contrast to the bottom-up scenario
alone, the hard top-down spectrum is able to produce a
pronounced recovery in the form of a flattening
beyond the ``cutoff'' which could explain the highest energy
events and possibly even a gap.

\section{Likelihood Analysis}
For the statistical analysis we assume that the data are
represented as the number of observed events, $n_i$, within a
given energy bin $i$, where $i=1,\cdots,N$. A given model
predicts a certain observed differential flux $j(E)$ (in units
of particles per unit area, unit time, unit solid angle and unit
energy). For this model the number of expected events, $\mu_i$, in
energy bin $i$ is then given by
\begin{equation}
  \mu_i=\int_{E_i^{\rm min}}^{E_i^{\rm max}}dE\,j(E)A(E)\,,
  \label{mui}
\end{equation}
where $A(E)$
is the total exposure of the experiment at energy $E$ (in units
of area times solid angle times time) and bin $i$ spans the
energy interval $\left[E_i^{\rm min},E_i^{\rm max}\right]$.
Both the Fly's Eye and the Akeno experiment used equidistant bins in
logarithmic energy space with log$_{10}(E_i^{\rm max}/E_i^{\rm
min})=0.1$. The likelihood function adequate for the low
statistics problem at hand is then given by Poisson statistics as
\begin{equation}
  L=\prod_{i=1}^N{\mu_i^{n_i}\over n_i!}\exp[-\mu_i]
  \,.\label{Ldef}
\end{equation}
Any free parameters of the theory are determined by maximizing
the likelihood Eq.~(\ref{Ldef}).
In analogy to Ref.~\cite{Agasa1} we then determine the
likelihood significance for the given theory
represented by the set of (optimal) $\mu_i$'s. It is defined as
the probability that this set of expectation values would by
chance produce data with a likelihood smaller than the
likelihood for the real data. This probability is
calculated by Monte Carlo simulation.

We will perform the fits in the energy range between
$10^{19}\eV$ and the highest energy observed in the respective
experiment. For comparison we compute the significance of these
fits in the range below the gap and in the range including the
gap and the highest energy events separately. This will
demonstrate the influence of this structure on the fit quality.

In determining the likelihood significance we also take the
finite experimental energy resolution into account. For the
statistical error we do that by folding the theoretical fluxes
with a Gaussian window function in
logarithmic energy space corresponding to an energy resolution
of about $30\%$. We determine the effect of the
systematic errors by repeating the procedure with data shifted
systematically by $\pm40\%$ in energy (see Tables~2 and 3).

Finally, for each model considered we simulate data for an
exposure increased by a factor $f$ assuming for this exposure
level the persistence of the apparent gap and the
flux associated with the highest energy events in
the existing data. This is done in the following way: For a given
model we determine the maximum likelihood fit to the real data as
described above which results in a set of expectation values
$\mu_i$ ($i=1,\cdots,N$). For all bins up to the second highest
energy observed we then draw random event numbers
$n_i^\prime$ from Poisson distributions whose mean values are
given by $\mu_i^\prime=(f-1)\mu_i$. This assumes that the
underlying model represents the data well below the gap (see
Tables 2) and continues to do so for increased exposure.
All other bins are assumed to contain no additional events,
$n_i^\prime=0$, except for the highest energy bin for which we
assume at least one more event, $n_i^\prime\geq1$.
The simulated data set then
consists of the sums of these numbers $n_i^\prime$ and the
numbers $n_i$ of events already counted. For this data set we
compute the likelihood significance of the underlying model as
above. By doing this many times one
can determine for the given exposure enhancement the confidence
level to which the given theory could be ruled out (or
supported) if the gap structure should persist.

\section{Input Models}
Before we present the results, let us describe the models we use
for $j(E)$. The Fly's Eye stereo data~\cite{Bird1,Bird2} show a
significant dip in the spectrum at around $5\times10^{18}\eV$.
In Ref.~\cite{Bird1} this was attributed to the superposition of
a steeper galactic component dominated by heavy nuclei and a
flatter extragalactic component of light particles like nucleons
or possibly also \g-rays. Above $\approx10^{19}\eV$
the latter one would thus dominate and a description by a
power law is consistent with the existing data at least up to
the GZK cutoff. For model 1 we therefore chose a power law
with normalization and power law index $q$ as free parameters.
A power law continuing beyond the GZK cutoff could be produced
in the following situations: First, there could be a nearby
source (i.e. nearer than a few $\Mpc$) of baryonic charged
particles for which the (power law) injection spectrum is not
noticeably modified and the GZK cutoff is irrelevant.
Second, the observed flux could be dominated by
neutral particles like \g-rays or even neutrinos~\cite{Halzen}
from a distant source~\cite{Sigl2}. Since in contrast to
nucleons there are no resonance effects in the interactions of
these particles around $10^{20}\eV$, their processed spectrum
would have a smooth shape which could be approximated by a power
law. The fits typically result in $q\approx2.7$ and thus
model 1 would belong to the bottom-up scenarios.

We also numerically calculated~\cite{Lee} the shape of the UHE CR
spectrum from single sources at various distances and from
uniformly distributed sources as it
would be observed after propagation through the intergalactic
medium. For all
these cases we used power law injection of primary
protons with cutoff energies $E_c\gg10^{21}\eV$, and
normalizations determined by maximizing the likelihood.
Our code accounts for the propagation of the nucleon component and
secondary \g-ray production as well as for \g-ray propagation.
Since current
experiments cannot distinguish between nucleons and a possible
\g-ray component, we used the sum of their fluxes for $j(E)$.
The secondary \g-ray flux depends somewhat on the
radio background and the extragalactic magnetic field
$B$~\cite{Lee}. In order to maximize the possible amount of
recovery we assumed a comparatively weak radio background with a
lower cutoff at 2MHz~\cite{Clark} and $B\ll10^{-10}$G. For
injection indices $q\geq2$ the resulting fluxes are
representative of acceleration models of UHE
CR origin. For the diffuse spectrum from a uniform source
distribution we assumed absence of source
evolution and chose $q=2.3$ which fits the data quite well below
the gap (see Table~2). The maximal source distance $d_{\rm
max}=10^3\Mpc$ was
chosen in a range where $d_{\rm max}$  has no significant
influence on the shape of the resulting spectrum above
$10^{19}\eV$. The minimal source distance $d_{\rm min}$ was
roughly chosen by maximizing the fit quality.

It turns out that a discrete source beyond a few $\Mpc$ alone
cannot explain the data
including the highest energy events. More interesting cases
are a diffuse spectrum alone (model 2), and its combination with
an additional discrete source at $10\Mpc$ (model 3), where in both
cases $d_{\rm min}=0$. In model 4 we combined a diffuse spectrum
for $d_{\rm min}=30\Mpc$ with a nearby source,
represented by an unprocessed power law with index
$q=2$. This model could be relevant if there were
a strong galactic source which accelerates iron nuclei much
beyond $10^{20}\eV$ with the hardest injection spectrum possible
for shock acceleration models ($q=2$).

Since in top-down models the flux above the GZK cutoff could be
dominated by \g-rays~\cite{Aharon} the processed
spectrum is somewhat uncertain due to interactions with
unknown backgrounds~\cite{Lee}. However, the hard top-down
component must be negligible below the GZK cutoff whereas above
the cutoff it can be approximated by a power law.
Similarly to model 4 we therefore chose best fit combinations of
the diffuse bottom-up spectrum for $d_{\rm min}=30\Mpc$ with an
unprocessed power law of index $q=0.6$ as our
model 5. Due to \g-ray propagation effects~\cite{Lee} the
injection spectrum corresponding to this latter hard component
could be considerably softer and consistent with various
constraints on energy injection as long as the X
particle mass is not too high~\cite{Sigl1,Wolf}.
Model 5 acts as a generic example of how
an additional hard top-down component might naturally
produce a pronounced recovery, i.e. a spectral flattening. In
fact we find that for $q<1$ the number of events expected per
logarithmic energy bin even starts to grow with energy beyond a
few times $10^{20}\eV$ although the actual differential flux is
always a decreasing function of energy. Indeed, this can
naturally give rise to a gap in the measured spectrum.

In Table~1 we summarize the main characteristics of the
representative models 1-5 discussed above.

\section{Results}
Table~2 presents the results for the Fly's Eye and Akeno data
available today for the models discussed in the previous
section. Below the gap a diffuse spectrum (model 2) is favored
by the data. Additional discrete sources beyond a few $\Mpc$
(model 3) do not improve the fit significantly. There is thus
no indication of a significant ``bump'' below the GZK cutoff
which would be produced by strong discrete sources~\cite{Hill1}.
If one includes the gap and the highest energy events into
consideration, an exclusive diffuse bottom up
component is ruled out at $90\%$ C.L. In contrast, bottom-up sources nearer
than a few $\Mpc$
are consistent with the data at the $1\sigma$ level.
Nevertheless, since there are no obvious visible
source candidates near the arrival directions of the highest
energy events observed, this is a highly problematic
option, as was argued in Ref.~\cite{Sigl}. Fig.~1A
shows the result of fitting the power law model 1 to the
Fly's Eye data from $10^{19}\eV$ up to the highest energy event.
The best fits in this energy range, however, are produced by
combinations of a
diffuse component with a hard unprocessed power law (models 4
and 5). Fig.~2A shows the result for the exotic model 5.

Table~3 summarizes the results obtained from the simulated
``data'' for a quadrupled exposure assuming that the gap and
the comparatively high flux in the highest energy bin persists
at this exposure. The constraints on
the models get much more stringent. Indeed, all curves
predicted from bottom-up models (model 1 to 4) can be ruled out at
least at the $98\%$ C.L., except
the most optimistic bottom-up model 4 involving a strong
nearby (supposedly iron) source, which could be ruled out only at
about $90\%$ C.L. The basic
reason is the following: Local sources can
reproduce the superhigh energy events, but at the same time
predict events in the gap which are not seen. Sources
beyond about $20\Mpc$ on the other hand predict a GZK cutoff
and a recovery which is much too weak to explain the highest
energy events. This conclusion can only be evaded by assuming a
systematic shift in observed energies of the order of
$40\%$ or larger. Fig.~1B shows a typical example
for fitting model 1 to data simulated for a quadrupled Fly's Eye
exposure.

In contrast, the representative model with a top-down component
is typically
consistent at the $1\sigma$ level as long as the shape of the
gap is not too discontinuous. Fig.~2B shows a
typical situation where the exotic model 5 is fitted to
simulated data. Thus, if the observed gap structure should
persist within a quadrupling of the data set it would be a
statistically significant proof of the need for new exotic
physics. We should stress that the significance for this would get
even more stringent if high fluxes would continue
considerably beyond the highest energies detected to date.
Conversely, if the gap structure should disappear and
the flux in the highest energy bins is not too high, there would
be no immediate need for new physics except for the non-trivial
problem of acceleration to such high energies~\cite{Norman}. A
decisive answer should definitely be possible with the proposed
Giant Air Shower
Array~\cite{Cronin} since it would allow enhancing the exposure
by much more than a factor 4. This instrument should also be
able to measure the composition of the UHE CR flux.

\section*{Acknowledgements}

We gratefully acknowledge James Cronin, Shigeru Yoshida and
Paul Sommers for reading the manuscript and giving valuable
suggestions.
This work was supported by the DoE, NSF and NASA at the
University of Chicago, by the DoE and by NASA through grant
NAG5-2788 at Fermilab, and by the Alexander-von-Humboldt
Foundation. S.L. acknowledges the support of the POSCO
Scholarship Foundation in Korea. P.B. wishes to thank Rocky Kolb
and Josh Frieman for hospitality and support at Fermilab at the
beginning stages of this work.

\newpage
\section*{Figure Captions}
\bigskip

\noindent{\bf Fig.~1.}
Maximum likelihood fits of the pure power law model 1 over the
energy range $10^{18.95}\eV\leq E\leq10^{20.55}\eV$.
We fitted the effective flux (dashed lines) which results from
the real differential flux (solid lines) by taking the
experimental finite energy resolution into account.
The data are given as $68\%$ C.L. error bars or as $84\%$ C.L.
upper limits. Note that for illustration purposes we multiplied
the steeply falling flux by $E^3$. ({\bf A})
shows the fit to the actual Fly's Eye monocular data and
corresponds to a likelihood significance of $46\%$ in the gap
region including the highest energy event. ({\bf B})
presents a typical example resulting from ``data''
simulated for an exposure enhanced by a factor 4 as described in
section~2. For these data model 1 would be ruled out at the
$98\%$ C.L.

\medskip
\noindent{\bf Fig.~2.}
Same as Fig.~1 but for the exotic model 5. In contrast to the
pure power law model 1 the likelihood significance in the gap region
($87\%$ and $47\%$ in case of ({\bf A}) and ({\bf B}), respectively)
typically stays within the $1\sigma$
level for both exposures.

\newpage
\renewcommand{\arraystretch}{2}
\tabcolsep0.5cm
\noindent{\bf Table~1.}
Summary of models used for the fits to the data. The models
consist of uniformly distributed sources (diffuse component), a
single source (discrete component) or a combination of these. We
give the source distance or range of source distances $d$ and
the power law injection index $q$. For a discrete source at
$d=0$ the power law injection spectrum is unmodified. The
normalizations of the components are fitted to the data.

\begin{table}[h]
\begin{tabular}{ccc}\hline
   & diffuse component & discrete component \\ \hline
   model 1 & $-$ & $d=0$, $q$ fitted \\
   model 2 & $0\leq d\leq10^3\Mpc$, $q=2.3$ & $-$ \\
   model 3 & $0\leq d\leq10^3\Mpc$, $q=2.3$ & $d=10\Mpc$,
   $q=2.3$ \\
   model 4 & $30\Mpc\leq d\leq10^3\Mpc$, $q=2.3$ & $d=0$,
   $q=2.0$ \\
   model 5 & $30\Mpc\leq d\leq10^3\Mpc$, $q=2.3$
   & $d=0$, $q=0.6$ \\ \hline
\end{tabular}
\end{table}

\newpage
\noindent{\bf Table~2.}
Likelihood significances for fits of various models to the
experimental data. The first number in each model row is for the
Fly's Eye monocular data and the second number is for the Akeno
data.
The fits were performed between $10^{19}\eV$ and the bin
containing the highest energy observed, corresponding to $E_{\rm
max}=10^{20.55}\eV$ and $E_{\rm max}=10^{20.4}\eV$, respectively.
Significances are given for the energy range below and above the
second highest event separately (left and right part). We used the
best experimental energy estimate (``central'') as well as
energies shifted systematically by $\pm40\%$.

\begin{table}[h]
\begin{tabular}{ccccccc}\hline
   & \multicolumn{3}{c}{$[10^{19}\eV-10^{19.9}\eV]$} &
     \multicolumn{3}{c}{$[10^{19.9}\eV-E_{\rm max}]$}\\
   & central & $-40\%$ & $+40\%$ & central & $-40\%$ & $+40\%$ \\
  \hline
  model 1 & 0.58 & $-$ & $-$ & 0.46 & $-$ & $-$ \\
          & 0.59 & $-$ & $-$ & 0.39 & $-$ & $-$ \\
  model 2 & 0.78 & 0.39 & 0.85 & 0.12 & 0.12 & 0.094 \\
          & 0.51 & 0.48 & 0.35 & 0.094 & 0.13 & 0.05 \\
  model 3 & 0.81 & 0.38 & 0.91 & 0.10 & 0.13 & 0.16 \\
          & 0.52 & 0.49 & 0.47 & 0.19 & 0.25 & 0.20 \\
  model 4 & 0.75 & 0.34 & 0.88 & 0.49 & 0.35 & 0.67 \\
          & 0.46 & 0.43 & 0.42 & 0.54 & 0.45 & 0.67 \\
  model 5 & 0.81 & 0.37 & 0.79 & 0.87 & 0.73 & 0.95 \\
          & 0.57 & 0.48 & 0.45 & 0.87 & 0.74 & 0.94 \\ \hline
\end{tabular}
\end{table}

\newpage
\noindent{\bf Table~3.}
Same as for Table~2 but using ``simulated data'' for a
quadrupling of experimental exposures assuming persistence of
the gap structure (see section 2). The likelihood significances are here
only given for the energy range of the gap including the
highest energy observed. ``$1\sigma$''
indicates that the model typically agrees with the simulated
data within the $1\sigma$ level as long as the gap
is not too discontinuous. Note that the Akeno sample
is in general less restrictive for the bottom-up scenarios since
it corresponds to a smaller exposure than the Fly's Eye sample.

\begin{table}[h]
\begin{tabular}{cccc}\hline
   & \multicolumn{3}{c}{$[10^{19.9}\eV-E_{\rm max}]$} \\
   & central & $-40\%$ & $+40\%$ \\ \hline
  model 1 & $\la0.02$ & $-$ & $-$ \\
          & $\la0.12$ & $-$ & $-$ \\
  model 2 & $\la0.02$ & $\la0.005$ & $\la0.05$ \\
          & $\la0.02$ & $\la0.02$ & $\la0.04$ \\
  model 3 & $\la0.02$ & $\la0.003$ & $\la0.05$ \\
          & $\la0.06$ & $\la0.02$ & $\la0.07$ \\
  model 4 & $\la0.15$ & $\la0.01$ & $\la0.3$ \\
          & $\la0.12$ & $\la0.03$ & $\la0.25$ \\
  model 5 & $1\sigma$ & $\la0.05$ & $1\sigma$ \\
          & $1\sigma$ & $\la0.11$ & $1\sigma$ \\ \hline
\end{tabular}
\end{table}

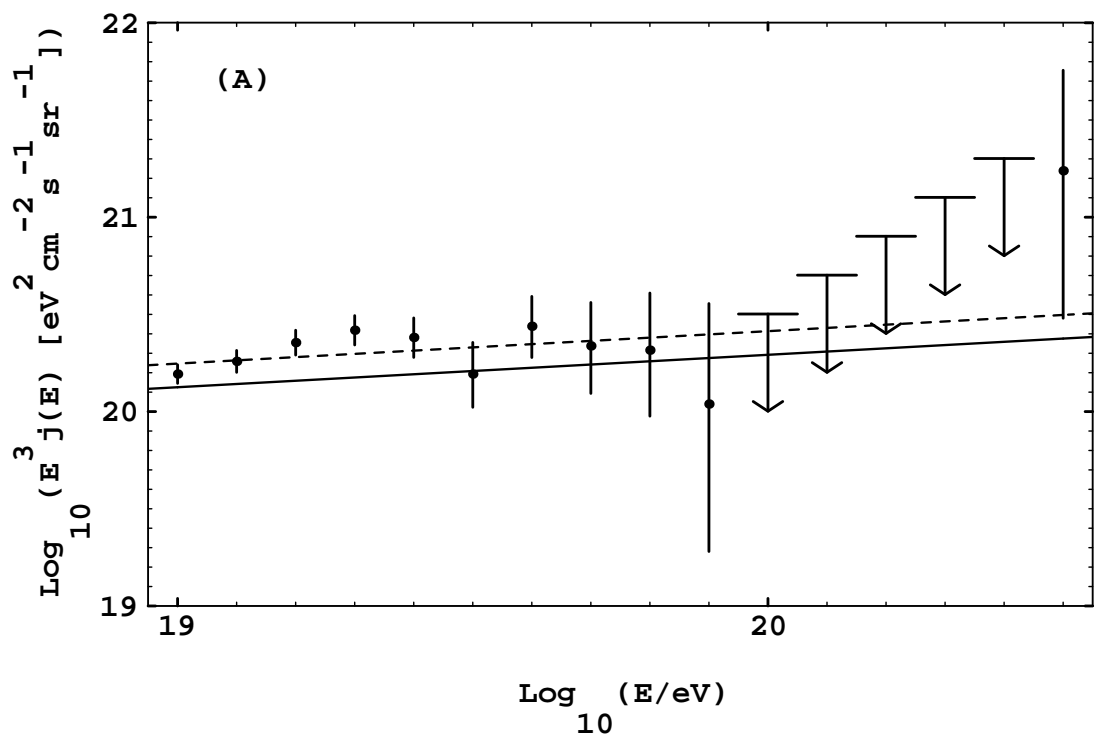
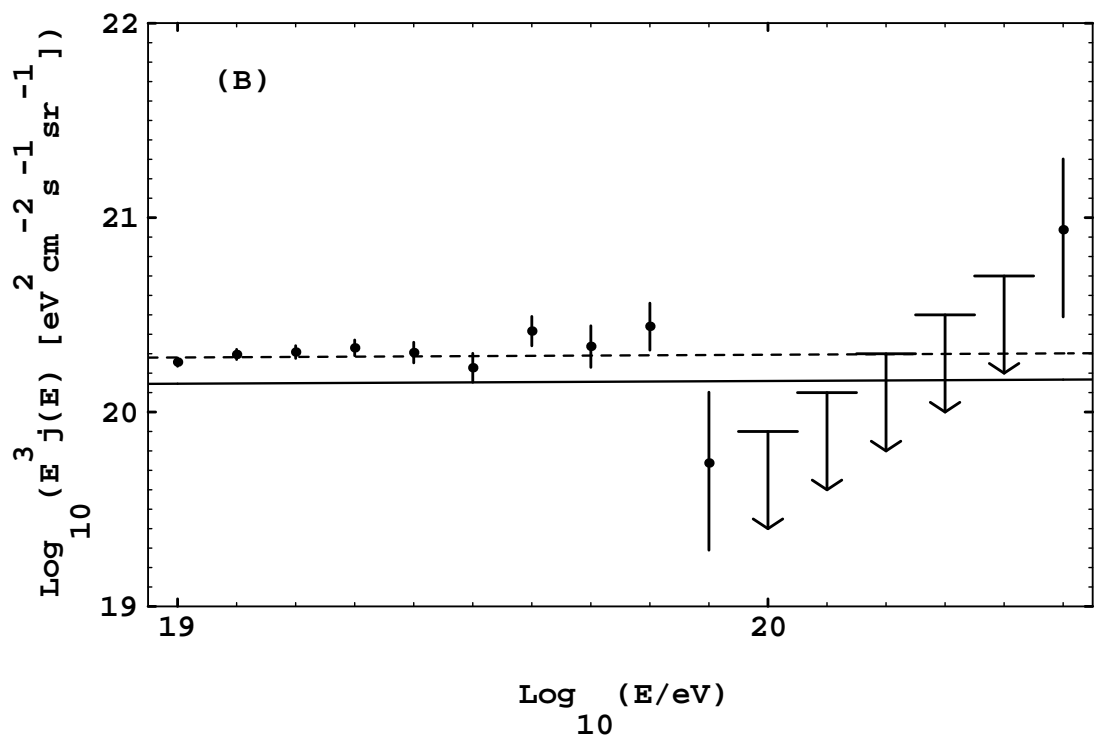
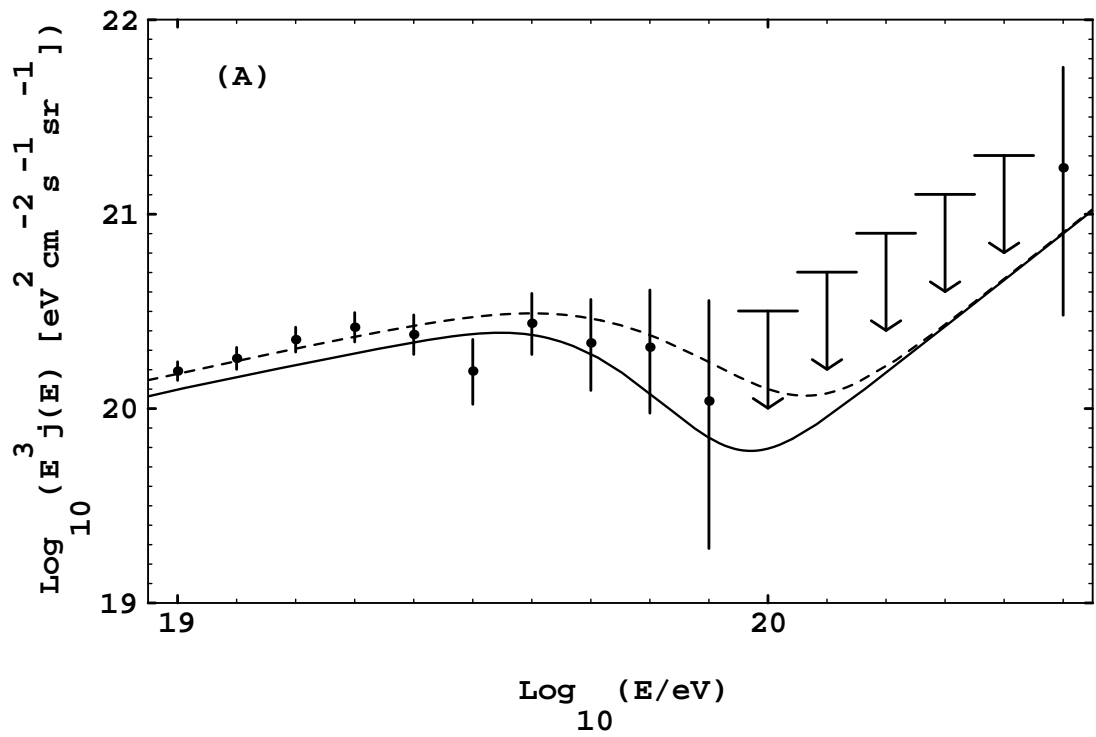
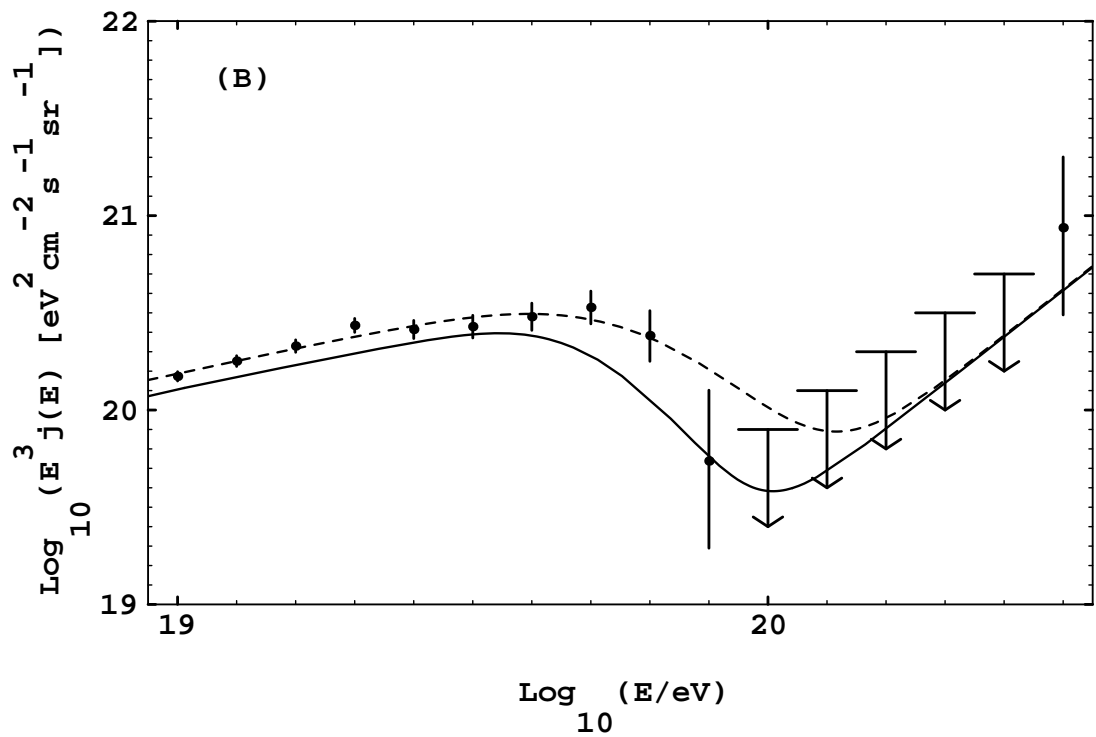
\end{document}